\documentclass[
   journal=aamick,
   manuscript=article,
   email=true,
   layout=traditional,
   amsmath,amssymb,linenumbers
]{achemso}

\usepackage{mciteplus}
\mciteErrorOnUnknownfalse

\usepackage[version=3]{mhchem} 
\usepackage{bm}
\usepackage{graphicx}
\usepackage{nicefrac}
\usepackage{booktabs}
\usepackage{xr}
\usepackage{xcolor}
\usepackage{hyperref}

\author{Francesco Delodovici}
\email{francesco.delodovici@centralesupelec.fr}

\author{Brahim Dkhil}
\affiliation{Université Paris-Saclay, CentraleSupélec, CNRS, Laboratoire SPMS, 91190, Gif-sur-Yvette, France}

\author{Charles Paillard}
\affiliation{Smart Ferroic Materials center and Institute for Nanoscience \& Engineering, Department of Physics, University of Arkansas, Fayetteville, Arkansas 72701, USA}
\email{ paillard@uark.edu}

\title{Size-effects on shift-current in layered CuInP\textsubscript{2}S\textsubscript{6}}

\begin{document}

\begin{abstract}
Two-dimensional ferroelectrics have recently emerged as a promising avenue for next-generation optoelectronic and photovoltaic devices. 
Due to the intrinsic absence of inversion symmetry, 2D ferroelectrics exhibit bulk photovoltaic effect (BPVE), which relies on hot, non-thermalized photo-excited carriers to generate a photo-induced current with enhanced performances thanks to efficient charge separation mechanisms. The absence of a required p-n junction architecture makes these materials particularly attractive for nanoscale energy harvesting.
Recent studies have reported enhanced BPVE in nanometer-thick CuInP$_2$S$_6$ ferroelectric embedded between two graphene wafers, driven by relatively strong polarization and reduced dimensionality. Short circuit photocurrent density values have been observed to reach up to $\rm mA/cm^2$.
In this paper, we demonstrate that the shift-current mechanism alone cannot fully account for these high conductivity values, suggesting that additional mechanisms may play a significant role.
Furthermore, our work confirms the existence of a strong size effect, which drastically reduces the shift-conductivity response in the bulk limit, in agreement with experimental observations.
\end{abstract}

\maketitle

\section{Introduction}

The bulk photovoltaic effect (BPVE)~\cite{Sturman1992} is a nonlinear optical phenomenon that enables highly efficient charge generation, potentially exceeding the Schockley–Queisser limit~\cite{Shockley1961, Spanier2016}. Unlike conventional photovoltaic mechanisms requiring p-n junction interfaces, BPVE does not require any built-in electric fields, as photocurrent is intrinsically generated in materials that break inversion symmetry. This makes BPVE particularly attractive for next-generation optoelectronic and energy-harvesting technologies.
Among materials hosting BPVE, two-dimensional van der Waals ferroelectrics \cite{Zhang_2023} have recently emerged as a promising platform for novel optoelectronics applications. 
Besides the theoretical interest for their rich physics \cite{Xiao_2022,Kim_2019,swamynadhan_2021,Liu_2016,Guan_2020}, their peculiar atomic arrangements made by strong covalent bonds in-plane and weak bonds out-of-plane, makes them CMOS-compatible and integrable with a vast range of materials, such as common semiconductor substrates such as silicon or gallium arsenide, with limited interfacial issues.
In particular in the field of photovoltaic energy conversion, they hold the potential for significant breakthroughs. 
On the one hand, they can be integrated in modern photovoltaic architectures, such as tandem-cells \cite{Zhang_2024,Li_2020}, to minimize recombination losses and improve overall efficiency.
On the other hand, they exhibit a BPVE with remarkably high performances \cite{Aftab_2022,Aftab_2023,Qiu_2023}, offering an alternative or complementary route to efficient charge generation beyond conventional semiconductor-based approaches.
It is the case, for instance, with CuInP\textsubscript{2}S\textsubscript{6}, recently recognized for hosting a large photocurrent density \cite{Li_2021}, up to $\rm mA/cm^2$, when embedded between two layers of graphene and illuminated by linearly polarized light.
However, current synthesis techniques seldom allow for the reproducible growth of perfect 2D ferroelectric monolayers on a large surface area, but rather isolated islands or films with a strong dispersion in thickness. It is thus very important to (1) quantify to what extent the BPVE, and in particular the shift current mechanism, contributes to the large photovoltaic response of CuInP\textsubscript{2}S\textsubscript{6}, (2) evaluate the role of sample thickness on the BPVE shift response, and determine whether there exists an optimal thickness yielding a maximal shift current output.
In this work, we investigate the origin of these large photocurrent responses using a Density Functional Theory based approach, which indicates that the shift-current mechanism alone is insufficient to explain the observed conductivity values. 
Our findings indicate that additional mechanisms contribute significantly, highlighting the need for further theoretical investigation. 
Furthermore, we confirm a pronounced size effect, where the shift-current response diminishes in the bulk limit, aligning with experimental observations \cite{Li_2021}. These insights provide a deeper understanding of BPVE in 2D ferroelectrics and its potential for future optoelectronic applications.

\section{System and methods}

CuInP\textsubscript{2}S\textsubscript{6} (CIPS) is a van der Waal layered material showing a ferroelectric phase transition aat around 315 K\cite{Maisonneuve_1997}. Following a second-order Jahn-Teller distortion coupling 3d$^{10}$ and 4s$^0$ orbitals in Cu and In, the resulting dipole ordering can be described as an antiparallel displacement of Cu$^+$ and In$^{3+}$ cations from the center of the surrounding sulfur framework. The asymmetry in the displacements of these ions results in a sizable reversible polarization of about 3.5 $\rm \mu C/cm^2$ at room temperature \cite{Maisonneuve_1997,Brehm_2020,Si_2019}.
We performed density functional theory (DFT) simulations with Quantum Espresso \cite{Giannozzi_2009}, using PAW pseudopotential \cite{blochl_1994} with PBEsol exchange-correlation functional~\cite{Perdew2008}. The convergence threshold for the self-consistency is set to $10^{-8}$ eV. 
We relaxed the bulk and few-layers slab configurations (1 to 4 layers, representing 20 to 80 atoms) until the forces are smaller than 10$^{-3}$ eV/\AA\, sampling the first Brillouin zone with a Monkhorst-Pack mesh having a density of at least 200 kpoint$\times$\AA$^3$.
In order to prevent the nonphysical interaction between periodic images, we truncated the Coulomb interaction in the z direction  \cite{Sohier_2017} and added at least 20 \AA\ of vacuum when simulating CIPS slabs.
The effect of semiempirical van der Waals corrections \cite{Grimme_2010} does not significantly affect the electronic properties of the system for the purpose of this study, as detailed in the Supplementary Information. Therefore, we do not include them in our simulations.
We calculate the shift current contribution to the BPVE and optical properties by post-processing maximally-localized Wannier orbitals~\cite{Azpiroz2018} obtained through Wannier90~\cite{Pizzi2020}.

\section{Results}
\subsection{Structure and electronic properties}

\begin{figure}
\begin{center}
 \includegraphics[scale=0.3]{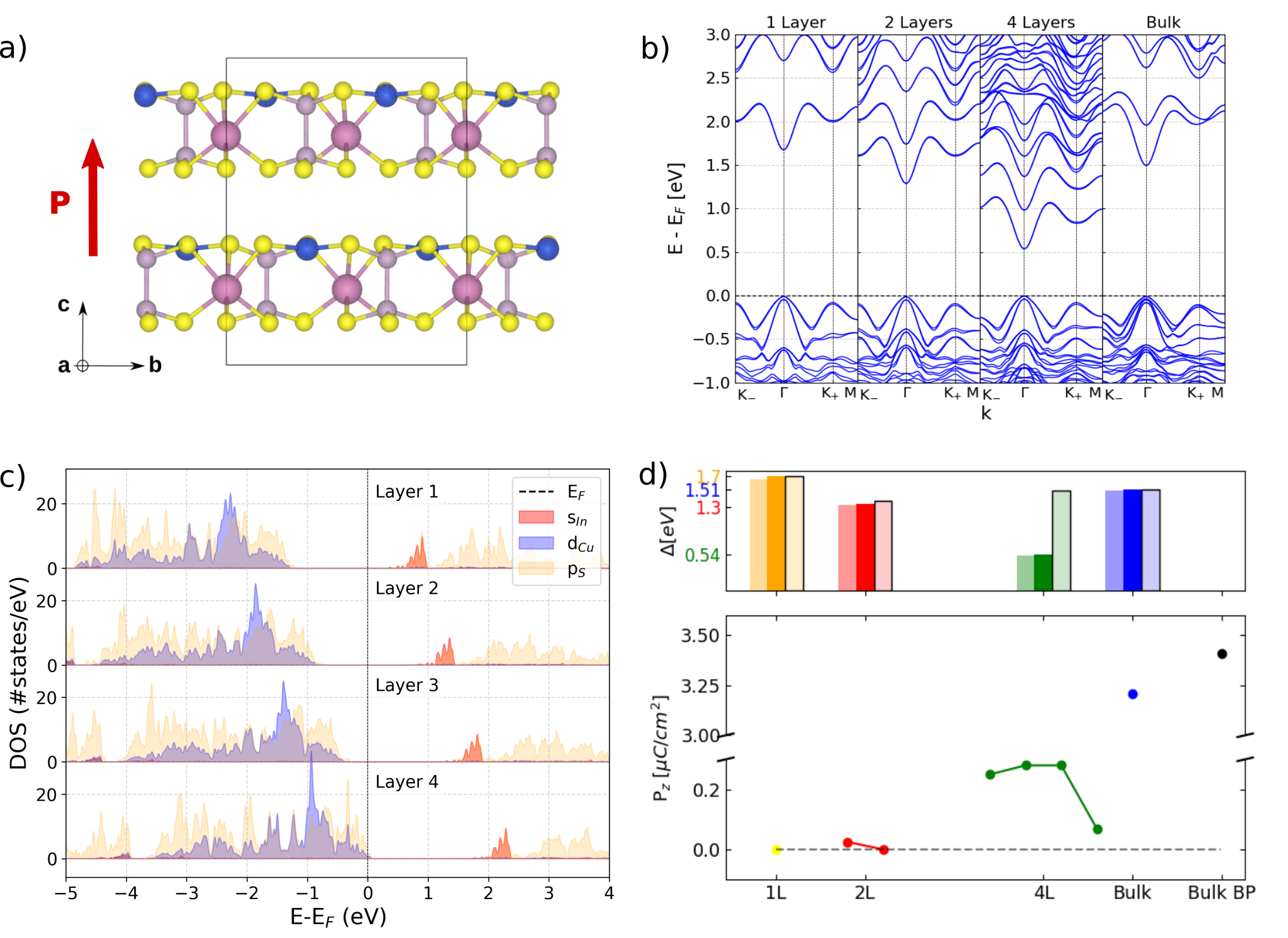}
 \caption{\textbf{a} CuInP$_2$S$_6$ monoclinic structure: Cu is reported in in blue, In in pink, P in grey, S in yellow. \textbf{b} The electronic band structure of the CIPS monolayer, bi-layer, four-layer, bulk configurations. The direct band gap is localized at $\Gamma$. \textbf{c} The atomic orbital projected DOS of the four layer structure. \textbf{d} Top panel: filled central bars represent the Homo-Lumo band gap: the colored labels report the corresponding values; black-border transparent bars represents the intra-layer averaged band gap: 1.34 eV and 1.48 eV respectively for the 2 and 4 layers slabs; transparent bars represents the optical gap extracted with Tauc analysis: 1.65 eV, 1.27 eV, 0.52 eV, 1.49 eV for 1-layer, 2-layer, 4-layers and bulk configurations. Bottom panel: the z component of the polarization as a function of slab thickness. Bulk phase P$_z$ obtained with Berry-phase is reported in black; with Born-effective charges in blue; layer-decomposed P$_z$: single-layers in yellow, bi-layer in red, 4 layers in green.}

 \label{fig:bands}
\end{center}
\end{figure}

The relaxed bulk primitive cell, reported in panel a of Fig. \ref{fig:bands}, is characterized by the monoclinic $Cc$ space group with primitive lattice parameters \textbf{a}=6.058\,\AA,\textbf{b}=10.490\,\AA,\textbf{c}=13.930\,\AA and $\beta$=107.12$^\circ$, and an inter-layer distance of 3.3 \AA. 
These values are in good agreement with experimental data at room temperature \cite{Maisonneuve_1997}, showing errors below 3\%. In contrast, the agreement with low-temperature values (around 100 K) is slightly less accurate \cite{Susner_2017}, in particular for the c lattice parameter, likely showing the increased relevance of van der Waals interactions in determining the structure at lower temperatures.
The bulk phase polarization, calculated using the Berry-phase method \cite{King-Smith_1993}, is 3.41 $\rm \mu C/cm^2$,  and it is almost parallel to the z-axis, with a minor in-plane component of P$_x = -0.19$ $\rm \mu C/cm^2$. The analysis of the polarization, obtained through the Born effective charges, as a function of the number of layers, reported in panel d of Fig.\ref{fig:bands}, shows how thickness reduction enhances the depolarization field, leading to a gradual, and eventually complete, suppression of P$_z$ and a strong reduction of $P_x$, as reported in Supplementary material. 
In parallel, the size reduction and corresponding enhanced electrostatic effects also influence the electronic properties, as can be seen in panels b, c and d of Fig. \ref{fig:bands}. The bulk material exhibits a direct band gap of 1.51 eV, with both the conduction band minimum and valence band maximum located at the $\Gamma$ point. As the thickness decreases, the band gap initially narrows due to the reduced screening of surface charges, inducing an electrostatic shift in the electron states energies. However, when approaching the monolayer limit, quantum confinement dominates, leading to a band gap larger than the bulk value \cite{Liu_2018}.
Panel c of Fig.\ref{fig:bands} clearly illustrates the effects of the depolarizing field, showing the energy shift in the orbital-projected density of states (DOS) within the four-layers slab. The dominant orbital character of the intralayer DOS, mainly indium s and sulfur p states near the band edges, remains consistent across all layers. The rigid energy shift is consistent with the direction of the polarization, pointing upward in the reported configuration.
The average intralayer band-gap, represented by the black-border bars in panel d of figure \ref{fig:bands}, confirms how the effect of the depolarizing field is larger in the 4 layer configuration, vanishing with P$_z$ when approaching thinner slabs.
To accurately capture the electronic features across different thicknesses, we performed non-collinear spin-polarized calculations including spin-orbit coupling at every stage. These simulations indicate the presence of a nontrivial spin texture in the valence bands near $\Gamma$, with more pronounced spin splitting emerging at the K$_{\pm}$ points, while the conduction bands remain less affected. More information can be found in the supplementary material. Although these spin features are noteworthy \cite{Zheng_2022,Picozzi_2014}, a detailed analysis lies beyond the scope of this work.

\subsection{Optical properties}

\begin{figure}
\begin{center}
 \includegraphics[scale=0.75]{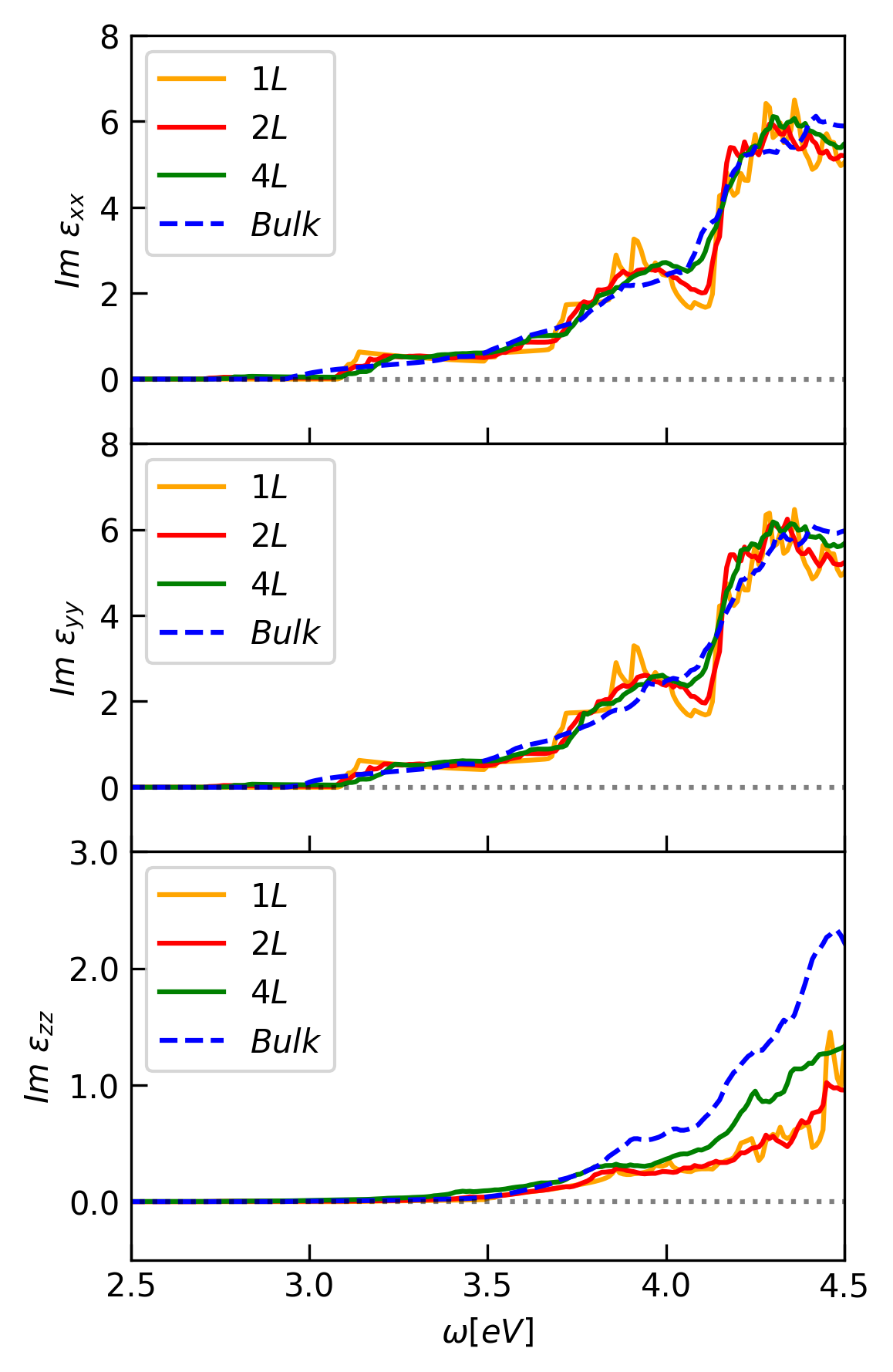}
 \caption{The imaginary diagonal components of the permittivity tensor for the single layer (yellow, top left panel), bi-layer (red, top right panel), 4-layers (green, bottom left), bulk (blue, bottom right) configurations. Different components are reported with different line styles.}
 \label{fig:epsilon}
\end{center}
\end{figure}

\begin{figure*}
\begin{center}
 \includegraphics[scale=0.85]{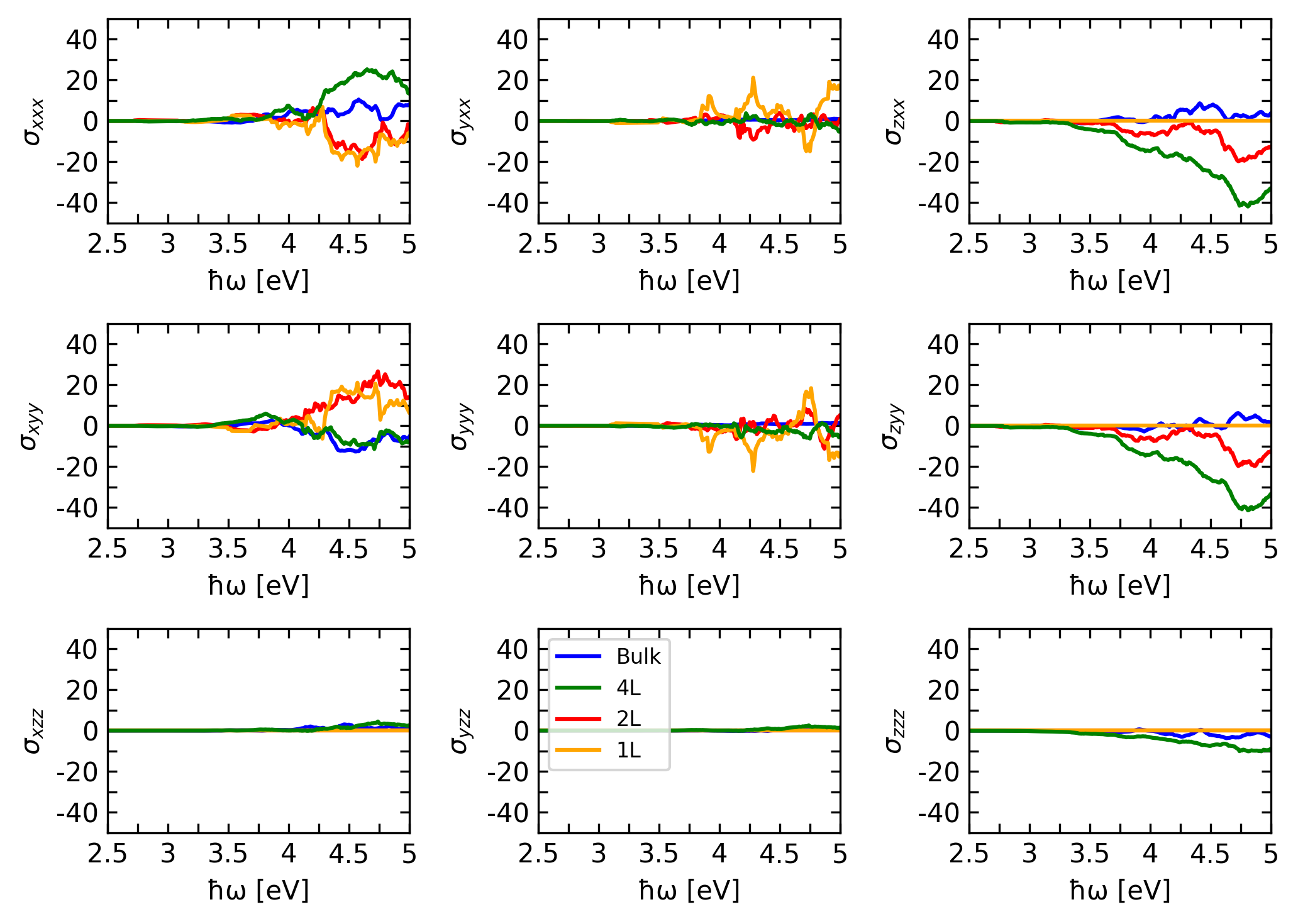}
 \caption{The shift current tensor components (in $\mu A \cdot V^{-2}$) corresponding to light linearly polarized along the lab reference frame axes for the bulk (blue line), 4-layers (green line), bi-layer (red line), mono layer (orange line) configurations.}
 \label{fig:shift}
\end{center}
\end{figure*}

\begin{figure*}
\centering
\includegraphics[scale=0.75]{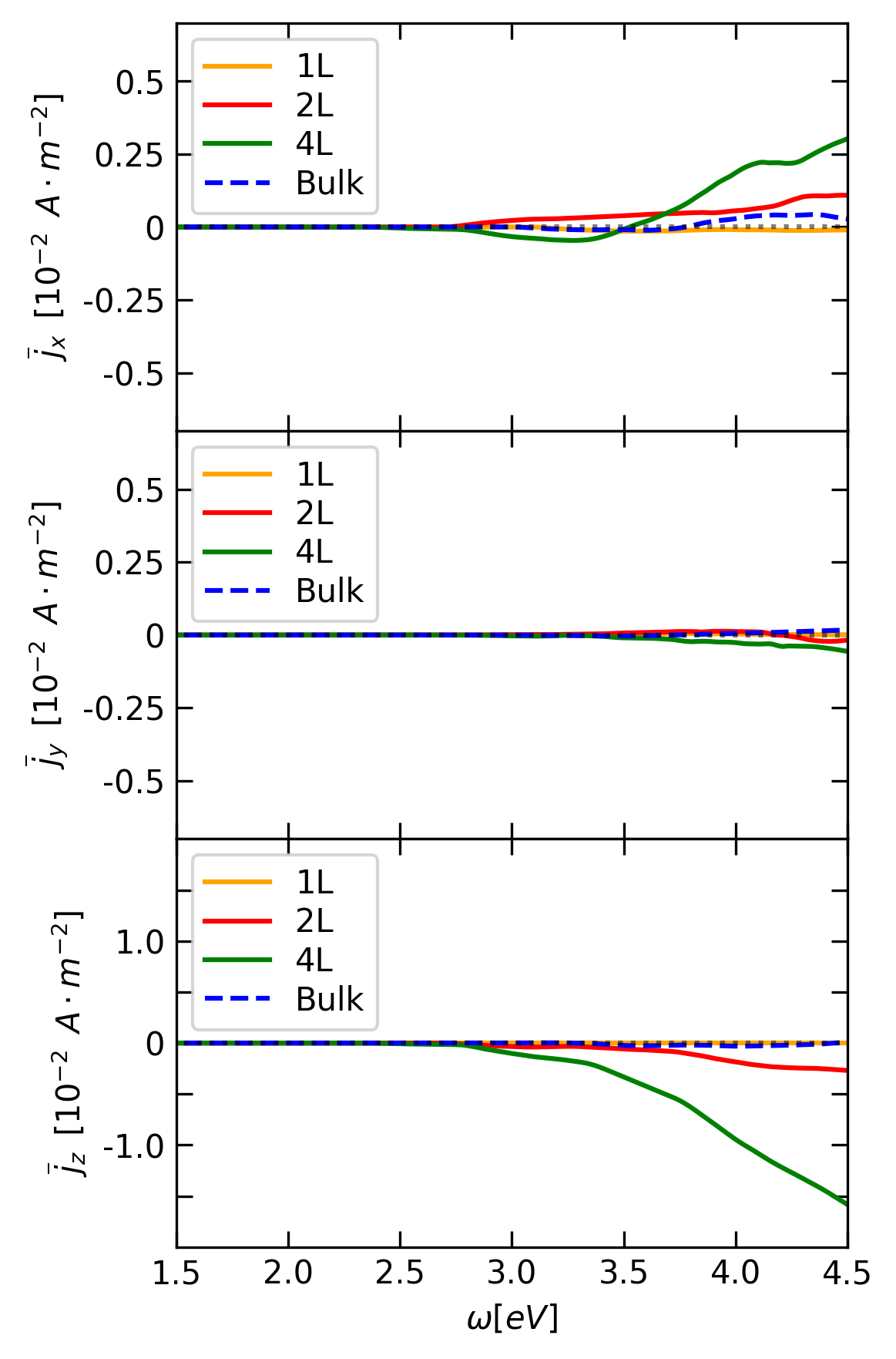}
 \caption{The weighted, and in-plane averaged, shift current density $\bar{j}_{k} = \frac{1}{2}(j_{ixx}+j_{iyy})$  along x (top panel), y (central panel) and z (bottom panel) directions, to the solar light spectrum when incident along z direction and penetrating a film of: 1 layer of CIPS, 6.7 nm thick (orange), 2 layers (red), 4 layer (green). The bulk case is reported in dashed blue as a comparison.}
 \label{fig:FOM}
\end{figure*}

Starting from the Kohn-Sham orbitals, we extracted the maximally-localized Wannier functions to compute the optical properties of various configurations. 
To recover the effective optical response of the 2D material, we rescaled the obtained numerical results as previously described in \cite{yang_2021,mu_2021,Jiang_2024,Azpiroz2018}, see supplementary material. 
We thus obtain: $\sigma''_{2D}=c/t\cdot \sigma''_{SB}$ and $\sigma'_{2D} = c/t(\sigma'_{SB}-1)+1 $, where $\sigma'$ and $\sigma''$ represent respectively the real and the imaginary part of the tensor, $\sigma_{SB}$ and $\sigma_{2D}$ indicate respectively the conductivity computed in the simulation box including vacuum, and the extrapolated 2D one. c and t represent, respectively, the simulation box thickness and the thickness of the slab. 
To remove the ambiguity in the definition of the 2D material thickness, we set $t$ as the center-to-center distance between the top and the bottom atomic layers in the bulk, multiplied by the number of layers. In our slab models, this averaged value varies by less than 2\%.

As the PBEsol functional systematically underestimates the band gap, yielding a value approximately $\approx 1.4$ eV lower than the experimental bulk-phase gap of $\approx 2.9$ eV \cite{studenyak_2003,Li_2021}, we apply a scissor shift to all calculated optical spectra to align them with the experimental band gap.
Figure \ref{fig:epsilon} shows the diagonal elements of the imaginary part of the permittivity as a function of light energy for the monolayer, bilayer, four-layer, and bulk configurations.
The spectral profiles remain overall qualitatively similar across different thicknesses, indicating that structural or electronic rearrangement upon thinning down to a few layers does not majorly affects optical properties.
Nevertheless, subtle changes emerge as the system approaches the bulk phase: the in-plane components ($\epsilon_{xx}$, $\epsilon_{yy}$) lose distinct features, such as the peak at 3.8 eV and the sharp onset near 3.2 eV, while the out-of-plane component ($\epsilon_{zz}$) shows a gradual increase with increasing thickness.
Additionally, owing to the surface states near the bottom of the conduction band in the thin-films, the absorption spectra exhibit characteristic low-energy tails, reported in supplementary materials, that are completely absent in the bulk profile.
Regardless of thickness, the systems display a pronounced anisotropic optical response. The out-of-plane component of the permittivity is consistently smaller than its in-plane counterpart. 
This anisotropy is further supported by the absorption coefficients given in the supplementary material. 
Despite its limitations for low-dimensional materials \cite{Klein_2023}, Tauc plot analysis \cite{Tauc_1966} can provide additional insights into the optical properties of the system, owing to the clearly direct character of the gap and to the well-defined parabolic behavior of the bands near the edges.
Our Tauc plot analysis, detailed in the supplementary material, confirms the presence of a slight anisotropy in the optical band gap.
This directional dependence reflects the layered structure of the material, indicating distinct contributions from interlayer and intralayer transitions to the optical absorption behavior, as evidenced for instance by the shifted density of states in Fig.\ref{fig:bands}.
The values reported in panel d of Fig.\ref{fig:bands} indicated by semi-transparent bars and representing the average optical band-gap, are in close agreement with the Homo-Lumo band gap estimate, within the fitting uncertainties (not shown for clarity). 

Let us now focus to the shift current tensor in CIPS. Figure \ref{fig:shift} displays the shift-current response $\sigma_{ikk}$ to linearly polarized light, in the case of mono-layer, bi-layer, four-layers and bulk CIPS. 
It is known that the shift current is larger for systems showing strong covalent bonds, highly delocalized orbitals~\cite{Rappe_2016,Tan_2019} and large anisotropy. In the case of van der Waals materials, due to their weakly interacting layered structure, the shift-current response to electric field polarized along the stacking direction is expected to be small, even vanishing. Our results confirm this interpretation, see e.g. $\sigma_{izz}$ in Fig.\ref{fig:shift}. 
Some weak contribution appears in $\sigma_{izz}$ for the 4-layers and the bulk configurations, likely produced by constructive interference of intralayer optical transitions.
Interestingly, different conductivity components do not evolve in the same way with respect to the thickness. 
For instance, the elements corresponding to out-of-plane transport produced by in-plane polarized field, i.e. $\sigma_{zii}$, increase with the number of layers well above the bulk value. This suggests the existence of an optimal number of layer maximizing the out-of-plane response of the system. On the other hand, the other components gradually evolve towards the bulk value.
To analyze the true solar-harvesting potential of CIPS in photovoltaic applications, we linked the shift conductivity to solar light absorption and to the film thickness by constructing an appropriate figure-of-merit $\bar{j}$.
We take the in-plane average of the conductivity weighted by the Planck distribution \cite{delodovici_2025} and by the light intensity profile as a function of in-film depth, and integrate over the light energy and on film thickness. More details can be found in the supplementary material. 
Notably, as reported in Figure \ref{fig:FOM}, while the integrated conductivity remains relatively flat in both the mono-layer and bulk configurations, with values barely exceeding a few $\rm 10^{-4} A\cdot m^{-2}$, the bi-layer and, in particular, the four-layer exhibit a markedly enhanced profile.
The behavior of a four-layers and bi-layers subjected to an in-plane oscillating electric field, represented by $\bar{j}_z$ , distinctly stands out from the rest, increasing monotonically from 3.25 eV onward and reaching $\rm 10^{-2} A/m^2$. 
This trend can be deduced from the behavior of the shift conductivity. The conductivity elements associated with transport along the stacking direction exhibit a constant sign as frequency varies, adding up constructively in the integration and producing the linear behavior reported in Figure \ref{fig:FOM}. In contrast, the pronounced oscillations in the $\sigma_{ykk}$ components prevents it from converging to a finite value when integrated. The behavior of $\sigma_{xkk}$ lies in between these two case.
Although they do not exhibit strong oscillations, $\sigma_{xxx}$ and $\sigma_{xyy}$ maintain opposite signs across the entire energy range considered. As a result, they interfere destructively when added.
As an alternative approach to evaluate the BPVE performances of few-layers CIPS, we computed the Glass coefficient \cite{Glass_1974,Rappe_2016} as the frequency-dependent ratio between the shift current and the absorbed light. 
The results, reported in supplementary, confirm that the few-layer configuration is best suited to effectively converts light into current via the shift-current mechanism.

\section{Discussion}

Our simulations confirm the strong dimensionality effect of BPV response in  CuInP$_{2}$S$_{6}$, in agreement with previous experimental findings.
In a real system, this intuitively stems from the fact that the thickness of the device, which corresponds approximately to the distance between the electrodes, is comparable to the carriers mean free-path, estimated  \cite{Li_2021}  to 40 nm in CIPS, which increases the transport efficiency.
In contrast, our work shows that this effect is also present in ideal systems, e.g. surrounded by vacuum, showing that this enhancement is probably also due to intrinsic effects \cite{cook_2017}, ultimately depending on the specificity of the bands structure rather than to the polarization magnitude, as reported in supplementary material. 
Nonetheless, our simulations show that, although it has a non-negligible contribution to the overall BPVE, the shift current is probably not the main contributor to the reported large photocurrent density~\cite{Li_2021}.
Indeed, the computed integrated shift-conductivity in the four/bi-layers is about one order of magnitude larger than the bulk one, in contrast to the three orders of magnitude increase reported in the literature.
The remarkable results observed for bilayer CIPS placed between two graphene electrodes may thus arise from interface-induced band bending, which leads to a conventional (Schottky interface-driven junction) photovoltaic effect rather than a bulk one.
To achieve a deeper comprehension and a better control of the BPV effect in CIPS, it would be important to investigate all contributions from the ballistic current too, including phonon-assisted mechanisms \cite{Dai_2021-phonon}, as well as those arising from quasiparticle \cite{Fei_2020} and excitons \cite{Dai_2021-el-hole} effects, although to-date, these effects have not proven to enhance the shift current by two orders of magnitude, to our knowledge. 
The relevance of the latter in 2D materials is still debated, even though it has proven to be of major importance in some recent work \cite{Esteve_2025,Lai_2024}.
Moreover, Cu$^+$ ion migration could also contribute through ionic conduction or charge accumulation at the CIPS-electrode interfaces.
Finally, as reported in supplementary material, we find that spin-orbit coupling substantially improves the photovoltaic response of the system, highlighting its critical influence on nonlinear optical properties. 
This suggests that SOC-driven effects may be leveraged to further optimize BPVE performances.
Finally, provided that spin transport is properly defined in SOC systems, few-layer CIPS may constitute a promising platform for investigating spin-BPVE and spintronics phenomena.

\section*{Acknowledgements}
C.P. acknowledges support from the Air Force Office of Scientific Research through Award No. FA9550-24-1-0263. F.D. and B.D. thank Agence Nationale de la Recherche for financial support through grant agreement no. ANR-23-CE09-0007 (SOFIANE) and n° ANR-24-CE08-0954-03 (PHOTOTRICS).

\bibliography{biblio}

\appendix
\section{Structural and electronic properties}

\begin{figure}[b]
\begin{center}
 \includegraphics[scale=0.65]{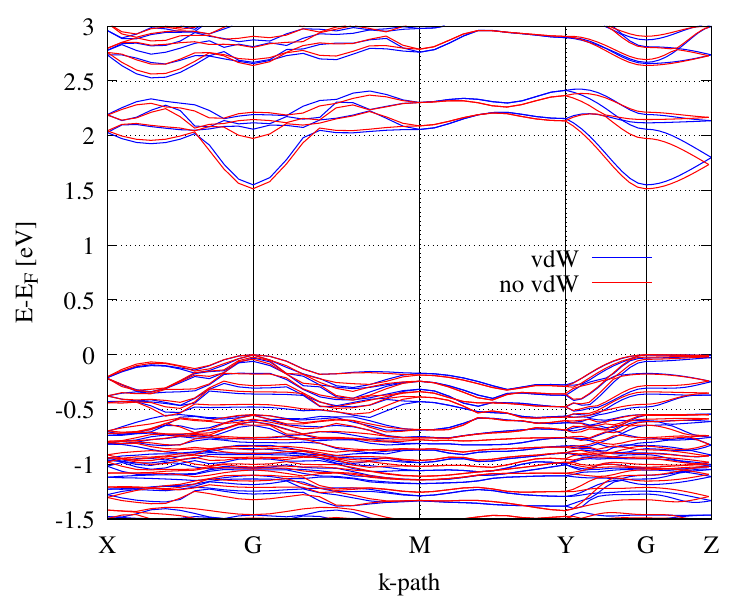}
 \caption{The electronic bands with and without vdW semi-empirical corrections in blue and red, respectively.}
 \label{fig:VDWbands}
\end{center}
\end{figure}

\begin{figure}
\begin{center}
 \includegraphics[scale=0.55]{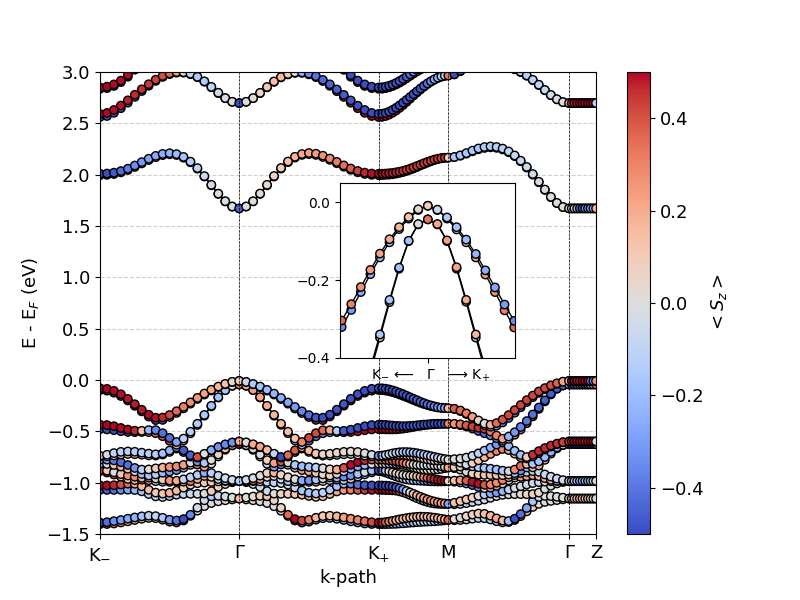}
 \caption{The spin-resolved electronic band structure of the CIPS mono layer. Red and blue colors represents the up and down spin components respectively. In the inset, a zoom around the $\Gamma$ point is reported to better show the effect of Rashba spin-momentum locking.}
 \label{fig:1Dbands}
\end{center}
\end{figure}

In table \ref{tab:structure} we report the effect of the van der Waals semi-empirical corrections on the lengths on the primitive cells, the angle $\beta$, the thickness t of the sulfur cage, and the electronic band gap of the bulk phase.
Figure \ref{fig:VDWbands} reports the effects of these corrections on the band structure of the bulk phase. The path is chosen in the monoclininc setting.
Figures \ref{fig:1Dbands}, \ref{fig:BLKbands} reports the spin-resolved band structure of mono-layer and bulk respectively.
The $\hat{S}_z$ expected value is represented by the color: 1/2 red , -1/2 blue.
The inset reports a zoom around $\Gamma$ to better appreciate the spin-texture forming due to broken inversion symmetry.
In this case, as in the bands reported in the main text, the path is taken in an equivalent setting, hexagonal in the x-y plane, \cite{Gjerding_2021}\cite{Curtarolo_2012}, defined: by a=b=6.06 \AA, c=13.46, $\alpha=\beta=94.17^\circ$,$\gamma=120^\circ$.

\begin{table*}[h]
    \centering
    \vspace{0.5cm}
     \caption{Effect of the van der Waals semi-empirical corrections on the
    primitive cell vectors, monoclinic angle $\beta$, sulfur-cage thickness $t$,
    and electronic band gap $\Delta$ in bulk CIPS.}
    \label{tab:structure}
    \begin{tabular}{c c c c c c c}
    \toprule
    \toprule
      & \textbf{a} [\AA] & \textbf{b} [\AA] & \textbf{c} [\AA] & $\beta\,[^\circ]$ & t [\AA] & $\Delta$ [eV] \\
    \midrule
    no vdW & 6.06 & 10.49 \quad & 13.93 & 107.12 & 3.08 & 1.52\\
    \addlinespace[2ex]
    Grimme & 6.03 & 10.44 & 13.48 & 107.42 & 3.3 & 1.55\\
    \bottomrule
    \bottomrule
    \end{tabular}
\end{table*}

\subsection{Polarization}

For the bulk phase the Born effective charges (BEC) underestimate the modulus of the polarization by less than 1\%: P is 3.41 $\rm \mu C/cm^2$ with Berry Phase (BP) method against 3.38 $\rm \mu C/cm^2$ with BEC. Nonetheless, effective charges tend to rotate the polarization towards the x axis: for the bulk phase P$^{BP}_{x}$=-0.19 against P$^{BEC}_{x}$=-1.06 , whereas P$^{BP}_{z}$=3.4 but P$^{BEC}_{z}$=3.21 $\rm \mu C/cm^2$.
The evolution of P$_z$ and P$_x$ with layer thickness is reported in Figure \ref{fig:Pz_Px}.
In the layer decomposition, we employed the Born effective charges computed for each slab configuration, and we normalized the polarization to the volume of the single bulk layers.
The z-component of P computed with the Berry-phase method for the bulk phase is reported as a comparison.

Figure \ref{fig:frozen_tau} compares the shift current calculated for bi- and four-layer relaxed structures with that obtained from corresponding configurations with frozen bulk ferroelectric displacements.
This comparison is aimed at assessing the sensitivity of the shift current to the magnitude of the polarization, and thus to the strength of the depolarizing field.
As evident from the figure, the dependence is rather weak: all tensor components largely preserve their shape and magnitude across the different configurations.
This supports the notion \cite{Rappe_2016} that the shift current response has a non-trivial relationship with polarization, that is not solely governed by its magnitude but is instead deeply connected to the topological features of the electronic band structure.

\section{Optical properties}

\subsection{Wannier orbitals}

Figure \ref{fig:wannier_bands} reports the bands computed with DFT as implemented in Quantum Espresso and those computed via Wannier interpolation with Wannier90 for the bulk phase.
The agreement is good over a range of 3-3.5 eV.
With the corresponding Wannier orbitals we computed the optical properties.

\subsection{Absorption spectra and Tauc plot}

The diagonal component of the absorption spectra are reported in figure \ref{fig:Abs}. The onset of the absorption spectra around the bottom of the valence band are reported in Figure \ref{fig:Abs_zoom}.
The absorption is obtained by processing the optical conductivity computed by postw90.x .

Figure \ref{fig:Tauc} reports the Tauc interpolation to extract the optical band gap from the absorption profile. The interpolation of $(\alpha \cdot h\nu)^{1/2}$ is performed over the energy windows $[2.93,2.96]\rm\,eV$, $[2.93,3]\rm\,eV$, and $[2.9,2.96] \rm\, eV$ for $\alpha_{xx}\,,\alpha_{yy}$ and $\alpha_{zz}$ respectively.
The Tauc analysis reveals the anisotropic nature of the system optical properties. The optical band gap obtained along x and y is respectively 2.92 ± 0.28 eV and 2.92 ± 0.2 eV in good agreement with experimental value $\approx2.9$ eV \cite{studenyak_2003,Li_2021}.
The extrapolation from $\alpha_{zz}$ lead a slightly smaller, but coherent, value: 2.89 ± 0.17 eV. This is likely due to the spatial separation of the valence and conduction bands. An example of this spatial separation is reported in Figure 1 of main text, even though in that case the spatial separation is enhanced by the presence of the depolarization field.

\subsection{Vacuum effects}

Figure \ref{fig:vacuum} reports the effects of different vacuum thickness in the simulation boxes, of the bi-layer slab, on some selected shift conductivity components corresponding to in-plane polarized light, for whom the effect is more evident.
In this case the thin film is electrically in parallel with the vacuum, thus in the static limit the conductivity will transform as equation 6 in \cite{Jiang_2024}, for optical fields it will transform as reported in the main text, see \cite{yang_2021}.

\subsection{Glass coefficient}

Figure \ref{fig:glass} reports the Glass coefficient under linearly polarized light G$_{ijj}(\omega) = \frac{1}{2c\epsilon_0} \frac{\sigma_{ijj}(\omega)}{\alpha_{jj}(\omega)}$ for the different studied configurations.

\section{Figure of merit}

The intensity of the electromagnetic field can be written as: 
$$ I(z) = \frac{\epsilon_0 c n'}{2} |E(z)|^2 = T\,I_0\,e^{-\alpha\cdot z}$$
where $\rm n'$ is the real part of the refraction index, $\rm c$ the speed of light, $\epsilon_0$ the vacuum permittivity, T the transmission coefficient from air to film, $\alpha$ the absorption coefficient (reported in Fig.\ref{fig:Abs}), $\rm I_0$ the light intensity at the film surface, and z represents the depth inside the film, taking as zero reference the top surface.
The shift current is then:
\begin{equation}
 j^i_{shift} (z) = \sigma_{ijk} \frac{2}{\epsilon_0 c n'} I(z) e_j e_k .
\end{equation}
Assuming normal incidence of the light ray, and considering the following Plank distribution as the light intensity $\rm I_0$:
\begin{equation}
    \tilde{B}(E,T) = \frac{1}{\hbar}B(\omega,T) 
\end{equation}
with dimensions of $\frac{W}{m^2\cdot J}$, we define the figure of merit as:
\begin{equation}
\begin{split}
    \bar{j}^{shift}_{i,k}&= \int_{0}^{\infty}\int_{0}^{t} \sigma_{ikk}(E) \frac{2}{\epsilon_0 c n'(E)}\,\tilde{B}(E,T)\,e^{-\alpha(E)\cdot z} T dz dE=\\
    \\
    &= \frac{2}{\epsilon_0 c}\int_{0}^{\infty}\frac{\sigma_{ikk}(E)}{\alpha_{zz}(E)} \frac{1}{ n'(E)}\,\tilde{B}(E,T)\,(1-e^{-\alpha(E)t}) dE =
    \\
    &= \frac{2}{\epsilon_0 c}\int_{0}^{\infty}\frac{\sigma_{ikk}(E)}{\alpha_{kk}(E)} \frac{1}{ n'(E)}\,\frac{1}{2\pi^2\hbar^3c^2}\frac{E^3}{e^{E/K_BT}-1}(1-e^{-\alpha(E)t}) dE\\
\end{split}    
\end{equation}
where we assumed a transmission coefficient T of 1. This object has the dimensions of $\bar{j}^{shift}=\frac{A}{m^2}$, so it correctly represents a current density.
The following quantity, averaged over x and y oscillating electric fields: 
\begin{equation}
    \bar{j}^{shift}_i = \frac{1}{2}(\bar{j}^{shift}_{i,x}+\bar{j}^{shift}_{i,y})
\end{equation}
corresponds to the quantity shown in Figure 4 in the main manuscript.

Figure \ref{fig:soc_effects} reports the effects obtained by introducing SO-coupling in the calculations.

\begin{figure}
\begin{center}
 \includegraphics[scale=0.55]{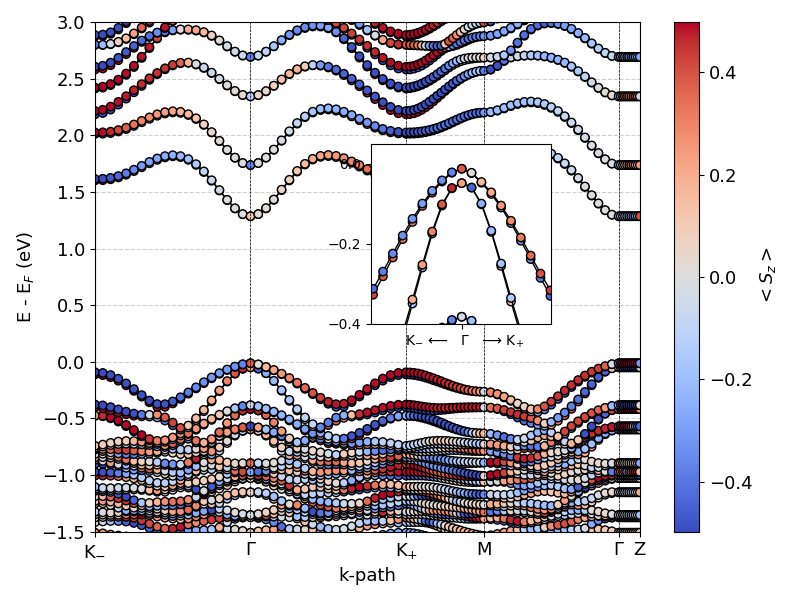}
 \caption{The spin-resolved electronic band structure of the CIPS bi layer. Red and blue colors represents the up and down spin components respectively. In the inset, a zoom around the $\Gamma$ point is reported to better show the effect of Rashba spin-momentum locking.}
 \label{fig:2Dbands}
\end{center}
\end{figure}

\begin{figure}
\begin{center}
 \includegraphics[scale=0.55]{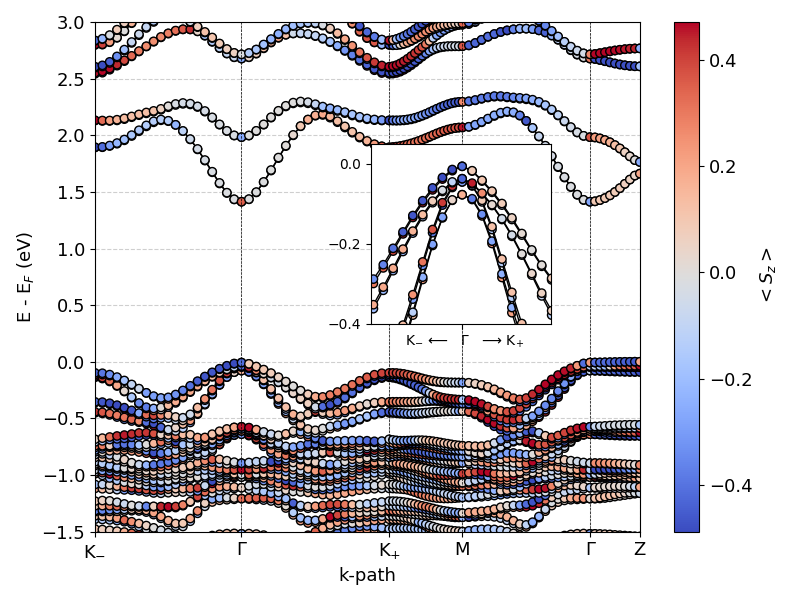}
 \caption{The spin-resolved electronic band structure of the CIPS bulk. Red and blue colors represents the up and down spin components respectively. In the inset, a zoom around the $\Gamma$ point is reported to better show the effect of Rashba spin-momentum locking.}
 \label{fig:BLKbands}
\end{center}
\end{figure}

\begin{figure*}
\begin{center}
    \includegraphics[scale=0.65]{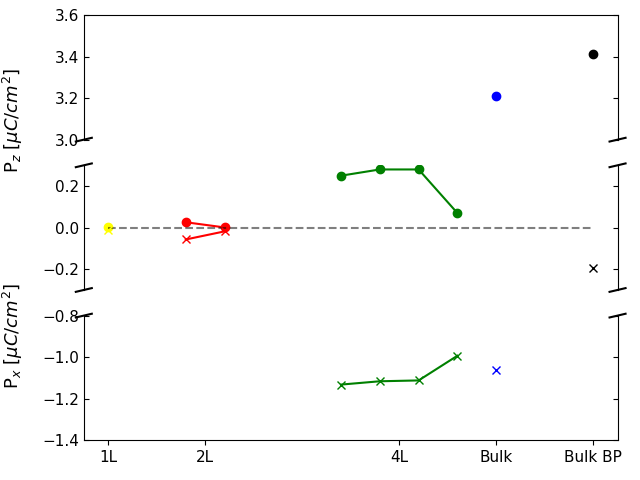}
    \caption{The z- and x-components of layer-projected polarization are reported.}
    \label{fig:Pz_Px}
\end{center}
\end{figure*}

\begin{figure*}
\begin{center}
 \includegraphics[scale=0.65]{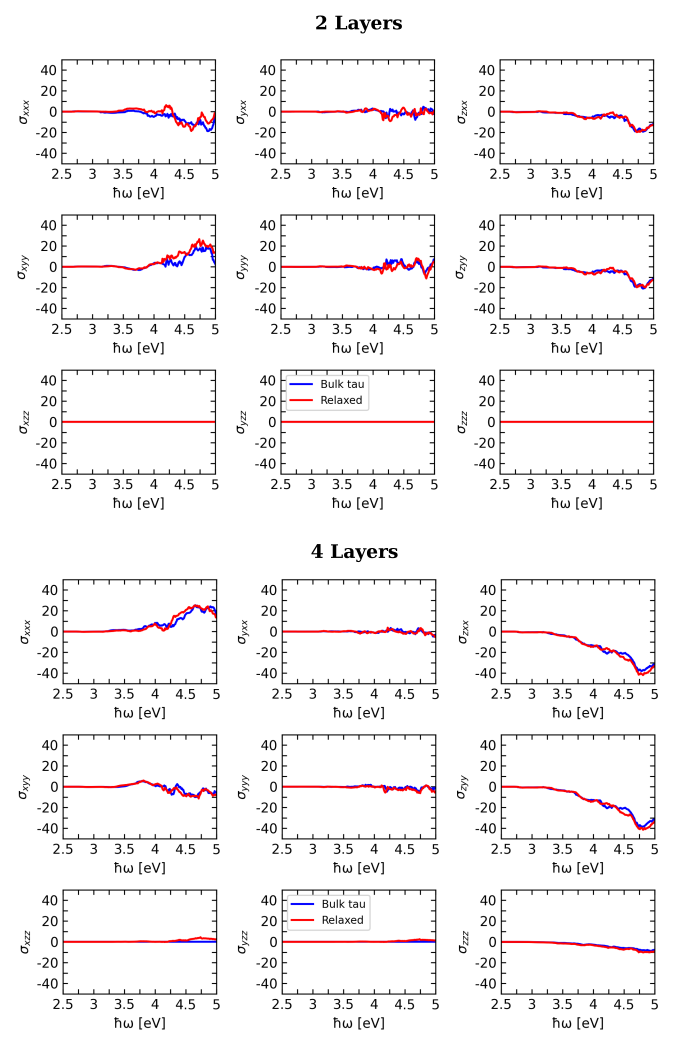}
 \caption{Top panel: relaxed bi-layer shift current (red) compared with shift current of bi-layer with ferroelectric displacements frozen to bulk value (blue). Bottom panel: relaxed four-layer shift current (red) compared with shift current of four-layer with ferroelectric displacements frozen to bulk value (blue).}
 \label{fig:frozen_tau}
\end{center}
\end{figure*}

\begin{figure*}
\begin{center}
 \includegraphics[scale=0.85]{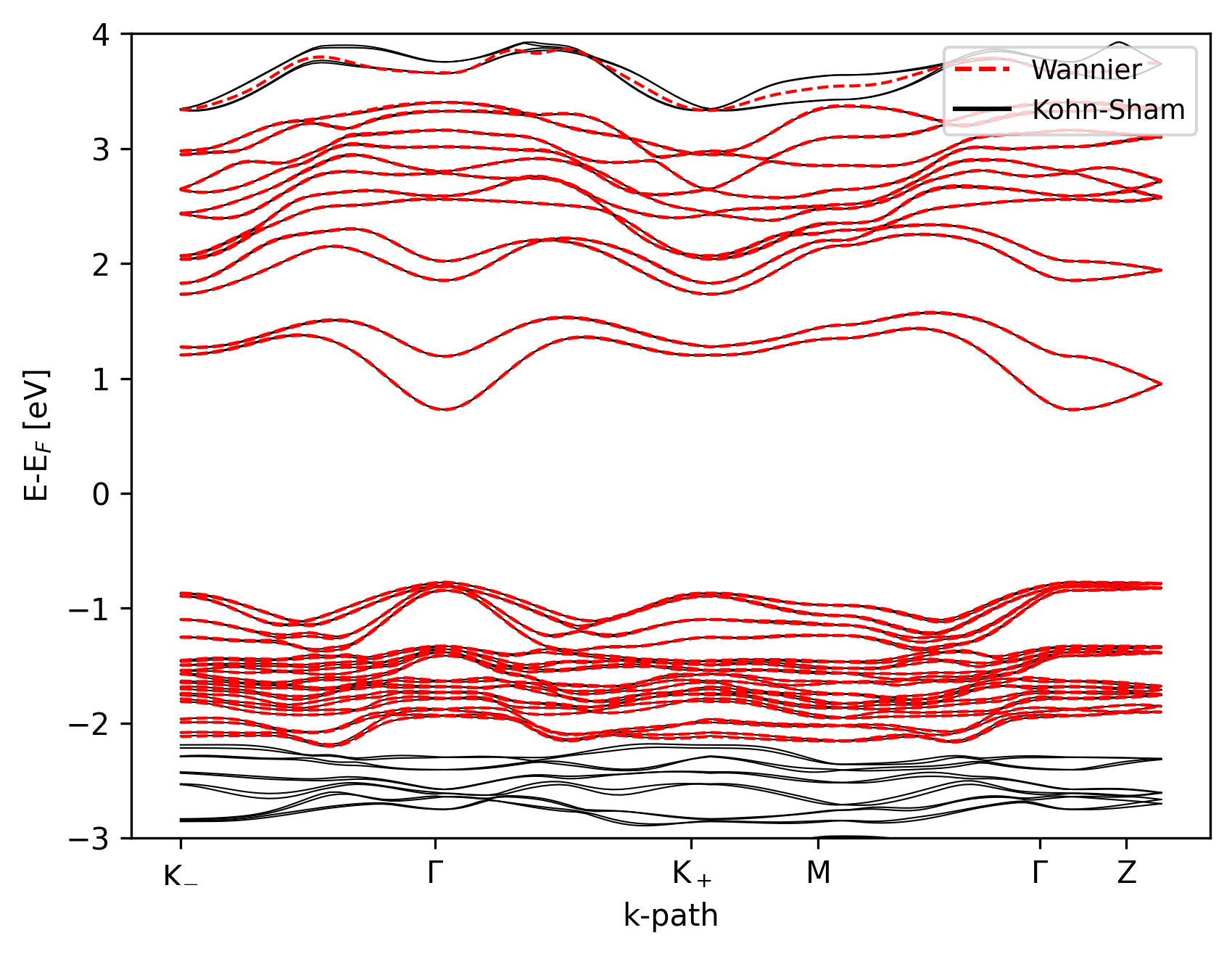}
 \caption{The electronic bands of the bulk configuration computed with DFT (black) and Wannier interpolation (red dashed).}
 \label{fig:wannier_bands}
\end{center}
\end{figure*}

\begin{figure*}
\begin{center}
 \includegraphics[scale=0.75]{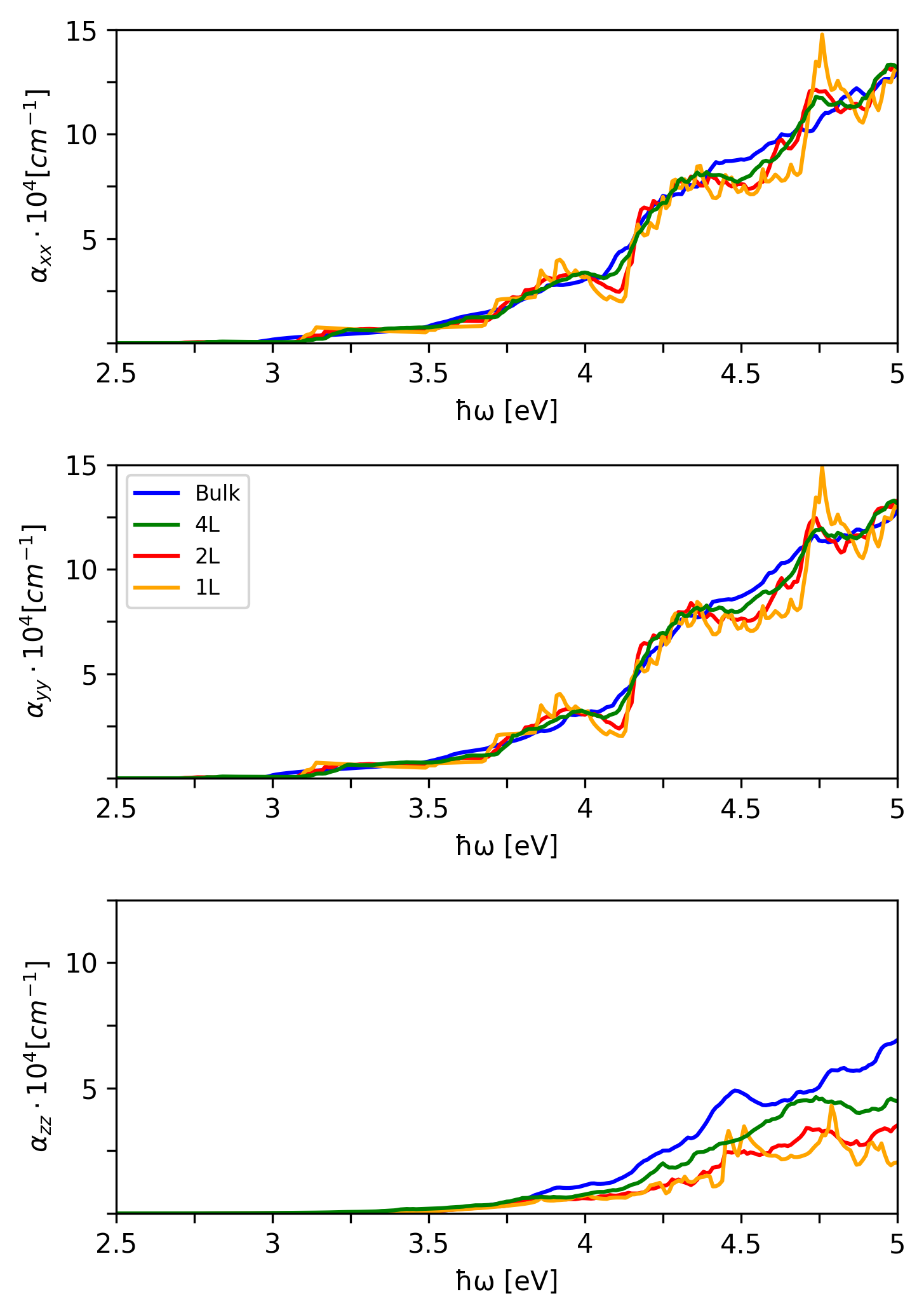}
 \caption{The diagonal components of the absorption tensor for bulk (blue line), four-layers (green), bi-layer (red), mono layer (yellow) configurations.}
 \label{fig:Abs}
\end{center}
\end{figure*}

\begin{figure*}
\begin{center}
 \includegraphics[scale=0.55]{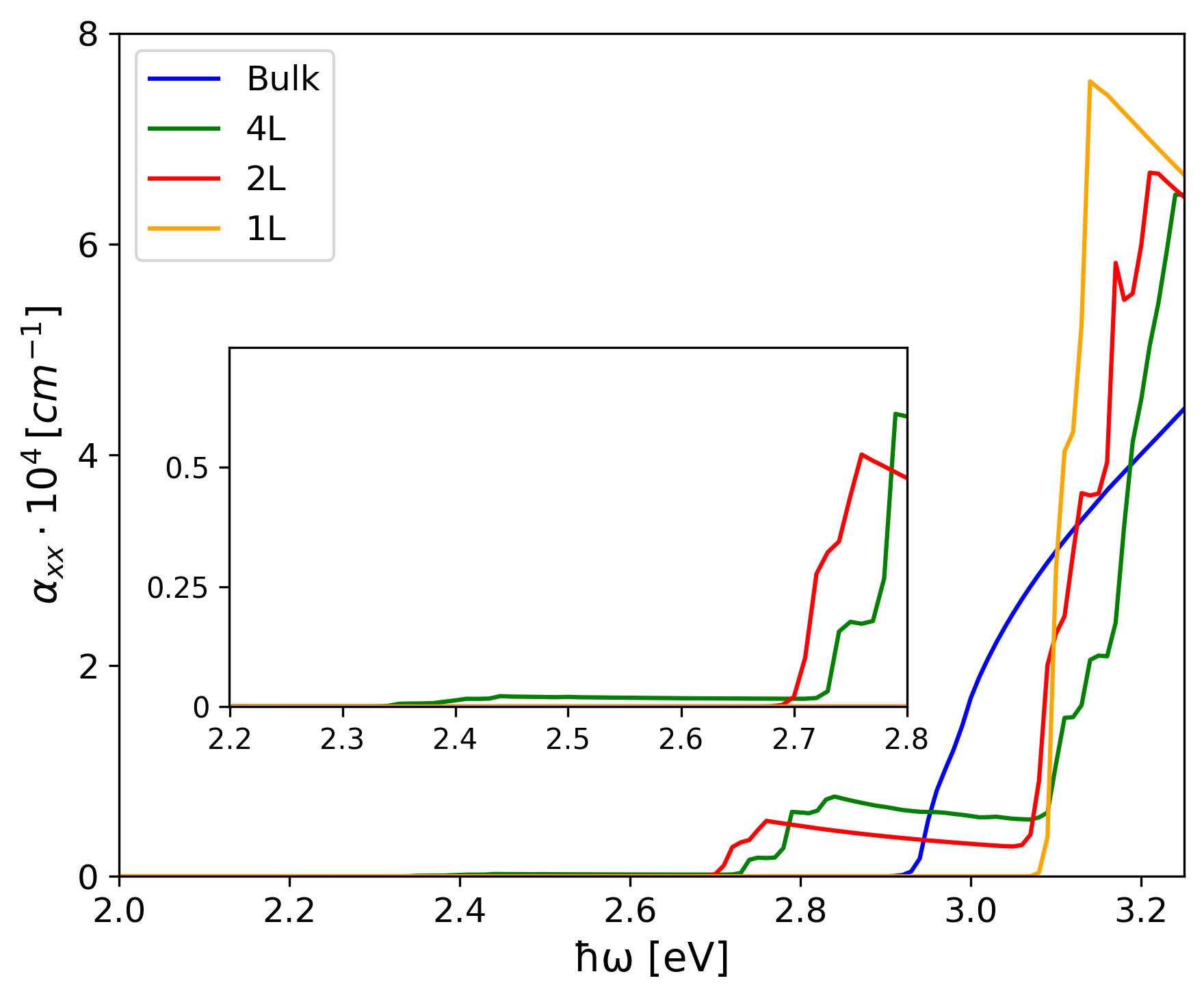}
 \caption{A zoom of the xx component of the absorption tensor around the minimum of the valence bands, for bulk (blue line), four-layers (green), bi-layer (red), mono layer (yellow) configurations. The inset shows the small contribution of the bottom of the conduction bands in the 4-layer slab.}
 \label{fig:Abs_zoom}
\end{center}
\end{figure*}

\begin{figure*}
\begin{center}
 \includegraphics[scale=0.65]{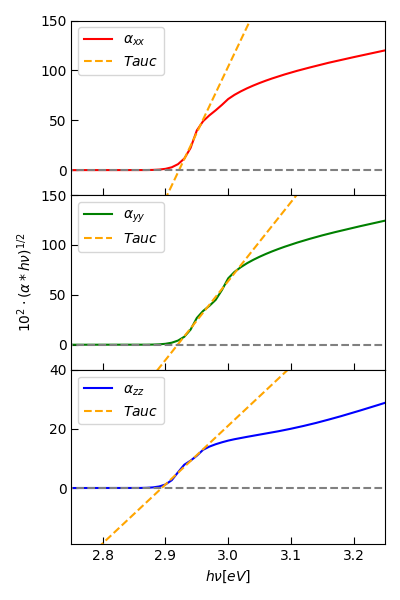}
 \caption{Tauc plot used to extract the optical band-gap for bulk configuration along the three cartesian directions.}
 \label{fig:Tauc}
\end{center}
\end{figure*}

\begin{figure*}
\begin{center}
 \includegraphics[scale=0.85]{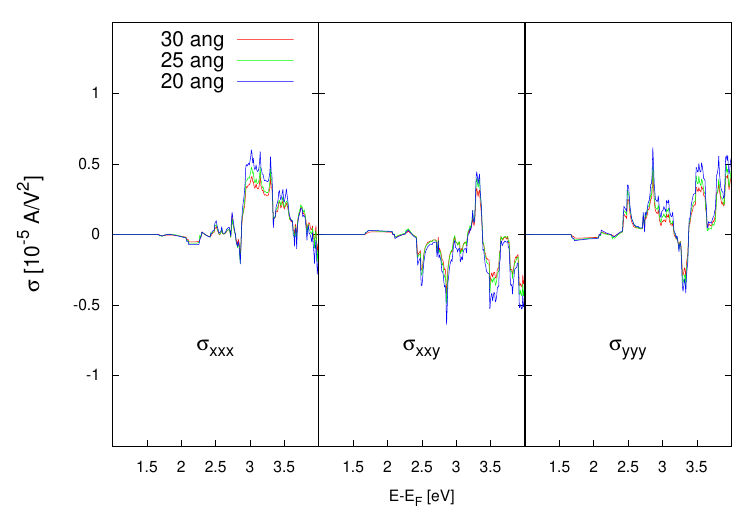}
 \caption{The effect of the simulation-box size on the shift current making rescaling necessary: 30 $\AA$ (red), 25 $\AA$ (green), 20 $\AA$ (blue) of vacuum.}
 \label{fig:vacuum}
\end{center}
\end{figure*}

\begin{figure*}
\begin{center}
 \includegraphics[scale=0.85]{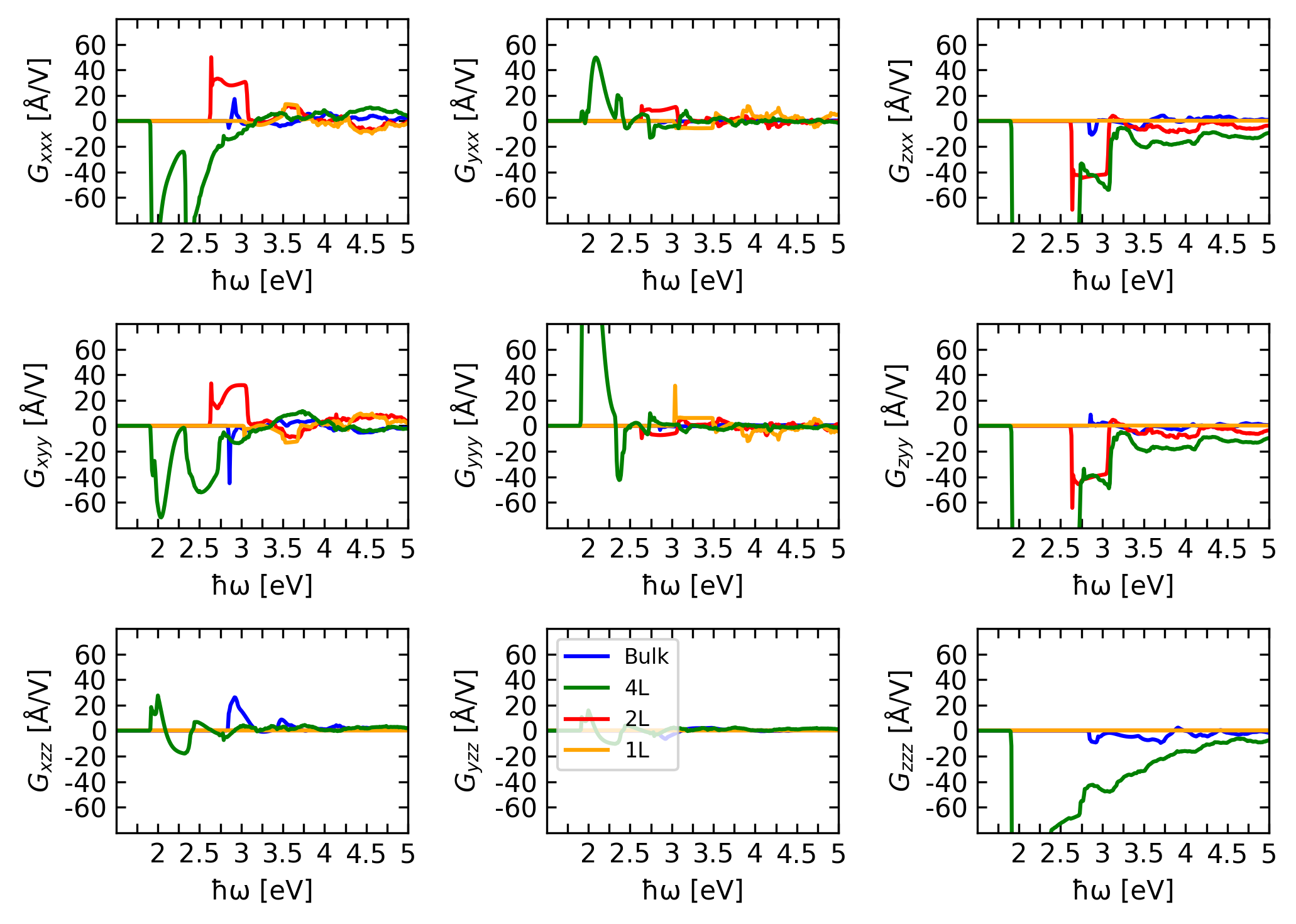}
 \caption{The glass coefficient for bulk (blue line), four-layers (green), bi-layer (red), mono layer (yellow) configurations}
 \label{fig:glass}
\end{center}
\end{figure*}

\begin{figure*}
\begin{center}
 \includegraphics[scale=0.65]{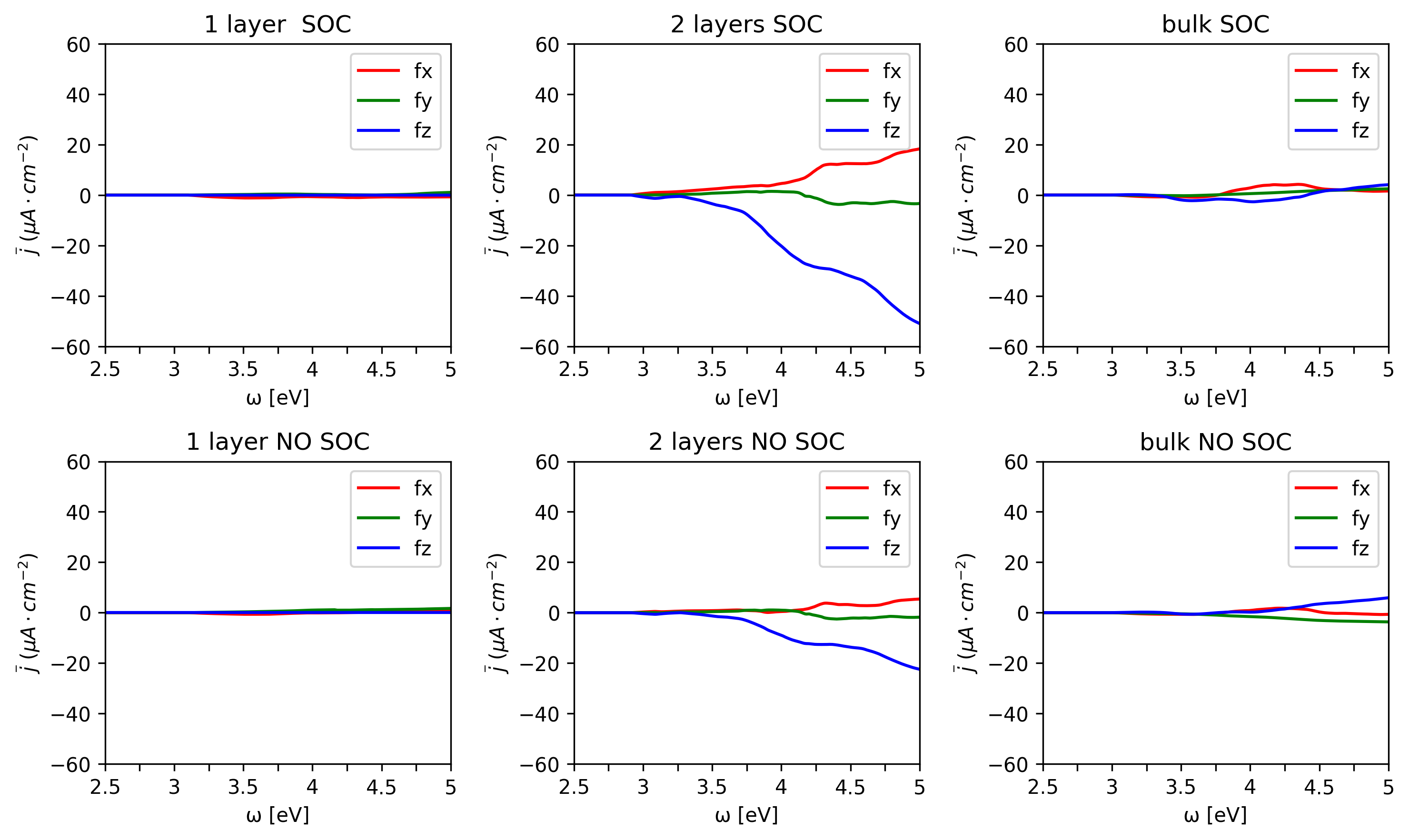}
 \caption{The effect of spin-orbit interaction on the proposed FOM: SOC included in the top panels, no SOC included in the bottom panels.}
 \label{fig:soc_effects}
\end{center}
\end{figure*}

%\bibliography{biblio}

\end{document}